\documentclass[final,5p,times]{elsarticle}
\usepackage{amssymb}
\usepackage{amsmath}

\usepackage{graphicx}

\journal{Journal of Magnetism and Magnetic Materials}

\begin{document}

\begin{frontmatter}

\title{Cation-ordered A$'_{1/2}$A$''_{1/2}$B$_2$X$_4$ magnetic spinels as magnetoelectrics}

\author{N.~V.~Ter-Oganessian}

\address{Institute of Physics, Southern Federal University, 194 Stachki pr., Rostov-on-Don, 344090 Russia}

\ead{nikita.teroganessian@gmail.com}

\begin{abstract}
We show that 1:1 ordering of A$'$ and A$''$ cations in A$'_{1/2}$A$''_{1/2}$B$_2$X$_4$ magnetic spinels results in appearance of magnetoelectric properties. Possible value of magnetically induced electric polarization is calculated using the recently proposed microscopic model, which takes into account spin-dependent electric dipole moments of magnetic ions located in noncentrosymmetric crystallographic positions. We build phenomenological models of magnetic phase transitions in cation-ordered spinels, which describe ferromagnetic and antiferromagnetic ordering patterns of B cation spins, and calculate the respective magnetoelectric responses. We find that magnetoelectric coefficients diverge at ferromagnetic or weak ferromagnetic phase transitions in ordered spinels.
\end{abstract}


\end{frontmatter}

\sloppy

\section{Introduction}

The search for new magnetoelectric materials has intensified in the last decade due to the discovery of whole new classes of magnetoelectrics and promising technological applications~\cite{Pyatakov_UFN_Review}. Multiferroic materials are sought among composite materials consisting of ferroelectric (piezoelectric) and magnetic (piezomagnetic) subsystems, as well as among single phase multiferroics. Single phase magnetoelectrics are usually divided into the so-called type-I and type-II multiferroics~\cite{Khomskii_Review}. In the type-I magnetoelectrics, or ferroelectromagnets, ferroelectricity and magnetic order occur independently and have different sources. Ferroelectromagnets (the prominent example is BiFeO$_3$~\cite{Catalan_Scott_Review}) usually possess high ferroelectric polarization, but the generally large difference between the ferroelectric and magnetic transition temperatures and the different causes of the two orders result in small coupling between them.

In contrast, the type-II magnetoelectrics, in which ferroelectricity occurs as a result of a magnetic phase transition, offer direct coupling between the magnetic and ferroelectric subsystems. Such magnetoelectrics, however, are characterized by much lower transition temperatures (10 -- 40~K) and low electric polarization values (usually of the order of 10 -- 100$~\mu\rm C/m^2$). The prominent examples of such magnetoelectrics are rare-earth manganites RMnO$_3$ (R=Gd, Tb, and Dy)~\cite{KimuraRMO3}. Cupric oxide CuO has the highest phase transition temperature ($\approx$230~K) among magnetoelectrics discovered to date~\cite{KimuraCuO}, which is still very low for practical applications. Therefore, the search for new single phase magnetoelectrics with higher phase transition temperatures is of paramount importance.

The spinel class with the general chemical formula AB$_2$X$_4$, where X=O, S, Se, or Te, and A and B are metals, is one of the richest structural classes~\cite{HMM_v3_ch4_oxide_spinels,HMM_v3_ch8_sulphospinels}. Spinels offer very high temperatures of magnetic phase transitions of the order of 1000~K (e.g. 860~K in Fe$_3$O$_4$, 1020~K in $\gamma$-Fe$_2$O$_3$, and 790~K in CoFe$_2$O$_4$)~\cite{HMM_v8_ch3_Ferrite_spinels} and allow for great flexibility in both cation and anion substitution~\cite{HMM_v3_ch4_oxide_spinels,HMM_v3_ch8_sulphospinels,HMM_v8_ch3_Ferrite_spinels}. All this makes spinels interesting from the point of view of searching for new materials and tailoring their properties.

The spinel compounds are widely used as constituents of multiferroic composites~\cite{Composites_Vaz}, whereas in single phase spinels the magnetoelectric effect (ME) has been found in a limited number of crystals. The ME effect was observed, for example, in CoCr$_2$O$_4$~\cite{CoCr2O4_ME_Yamasaki} and ZnCr$_2$Se$_4$~\cite{ZnCr2Se4_Murakawa}. However, these magnetoelectric spinels are characterized by low magnetic phase transition temperatures (of the order of 20~K) and incommensurately modulated magnetic order of the ferroelectric phase.

In this work we analyze magnetic spinels with the general chemical formula A$'_{1/2}$A$''_{1/2}$B$_2$X$_4$. We show that the chemical ordering of the A$'$ and A$''$ cations results in the appearance of magnetoelectricity in such spinels.

\section{Atomic ordering in A$'_{1/2}$A$''_{1/2}$B$_2$X$_4$ spinels}

The high symmetry AB$_2$X$_4$ cubic spinel structure is described by the space group Fd$\bar{3}$m ($O_h^7$). If more than one sort of cations is present in one of the equivalent sublattices, a tendency generally exists to decrease the internal energy by ordering the cations. Such ordering may possess both short and long range characters, depending on the energy gain and thermodynamic history of the crystal. As a general rule one may state that the higher the difference in valences of inequivalent cations the stronger their tendency to order~\cite{HMM_v3_ch4_oxide_spinels}.
The examples of spinels exhibiting atomic ordering in the A sublattice are Li$_{1/2}$Ga$_{1/2}$Cr$_2$O$_4$, Li$_{1/2}$In$_{1/2}$Cr$_2$O$_4$~\cite{LiGaCr2O4_LiInCr2O4_Okamoto}, Cu$_{1/2}$In$_{1/2}$Cr$_2$S$_4$~\cite{CuCr2S4_based_Kesler}, and Fe$_{1/2}$Cu$_{1/2}$Cr$_2$S$_4$~\cite{Fe0.5Cu0.5Cr2S4_Palmer,FeCuGaCr2S4_Aminov}, whereas the atomic ordering in the B sublattice may be attained in Zn[LiNb]O$_4$, Zn[LiSb]O$_4$, and Fe[Li$_{1/2}$Fe$_{3/2}$]O$_4$~\cite{HMM_v3_ch4_oxide_spinels}.

Various types of cation orderings in spinels (1:1 in the A sublattice, $\alpha$ and $\beta$ 1:1 ordering in the B sublattice,  1:3 ordering in the B sublattice, and others) are considered in~\cite{Ordering_Spinel_Haas,Ordering_Talanov} and possible orders of the order-disorder phase transitions are established. The atomic ordering results in loss of some symmetry elements, which reduces the cubic symmetry. In this work we focus on 1:1 cation ordering of A cations in A$'_{1/2}$A$''_{1/2}$B$_2$X$_4$ spinels, which results in every A$'$ cation surrounded by four A$''$ cations and viceversa. Such ordering leads to reduction of the crystal symmetry to F$\bar{4}$3m ($T_d^2$)~\cite{Ordering_Spinel_Haas} locally if only short range ordering is attained or globally if a long range order is established.

The distribution of atoms over the lattice sites measured by the degree of atomic ordering is an important property of multiatomic crystals. The atomic ordering degree depends on the thermodynamic history of the sample or synthesis conditions and frequently can be varied to a large extent. Among such crystals are ordering alloys and multiatomic compounds such as oxides and halogenides. If such crystals undergo structural or magnetic phase transitions, the temperatures of these transitions and the macroscopic properties of the crystals strongly depend on the type and degree of atomic ordering. Such behavior is, for example, ubiquitous in the perovskite class ABO$_3$, which offers great potential for ion substitution. For example, disordered and ordered samples of PbSc$_{1/2}$Ta$_{1/2}$O$_3$ show completely different dielectric behavior~\cite{PbScTaO3_Chu}, whereas cation ordering in SrFe$_{1/2}$Mo$_{1/2}$O$_3$ significantly influences the magnetotransport and magnetic properties~\cite{SrFeMoO3_Sarma,SrFeMoO3_Sanchez-Soria}.

When interpreting the influence of atomic ordering on properties of crystals one usually proceeds with the assumption that the degree of atomic ordering $s$ makes quantitative contribution to the thermodynamic potential~\cite{Wagner_Silin}. Within the framework of phenomenological theory this approach reduces to the introduction of the dependence on $s$ of the coefficients in the thermodynamic potential expansion with respect to the relevant order parameters~\cite{Bokov}. However, it was shown that the influence of atomic ordering can be much more substantial~\cite{Sakhnenko_2003,Sakhnenko_2005}. Namely, at $s\neq 0$ additional contributions to the thermodynamical potential may arise, which are forbidden by symmetry in the disordered case $s=0$. These contributions manifest themselves especially strong when they include degrees of freedom, which are described by macroscopic tensors. This results in formation of corresponding macroscopic fields during the phase transitions and divergencies in the corresponding susceptibilities~\cite{Sakhnenko_2003}.

The 1:1 cation ordering in the A sublattice of A$'_{1/2}$A$''_{1/2}$B$_2$X$_4$ spinels is described by the order parameter $s$ transforming according to the irreducible representation (IR) GM$^{2-}$ of the space group Fd$\bar{3}$m. Denoting by $N_{A'}$ and $N_{A''}$ the number of atoms $A'$ and $A''$, respectively, in one of the sublattices appearing upon atomic ordering, we can define the atomic ordering degree as
\[
s=\frac{N_{A'}-N_{A''}}{N_{A'}+N_{A''}}.
\]
Thus, $s$ varies from zero for completely disordered crystal to $\pm1$ for completely ordered one. Nonzero $s$ results in disappearance of center of inversion and lowering of the crystal symmetry to F$\bar{4}$3m. The emergence of noncentrosymmetric structure upon 1:1 cation ordering in A$'_{1/2}$A$''_{1/2}$B$_2$X$_4$ spinels is to be contrasted with 1:1 cation ordering in A$'_{1/2}$A$''_{1/2}$BO$_3$ or AB$'_{1/2}$B$''_{1/2}$O$_3$ perovskites, where such ordering results in centrosymmetric crystal lattice.

\section{Magnetic phase transitions in spinels and atomic ordering\label{sec:Magnetic_Phase_Transitions}}

Spinels exhibit a variety of magnetic structures including ferromagnetic (e.g. CuCrZrS$_4$~\cite{CuCrZrS4_Iijima}), ferrimagnetic (e.g. FeCr$_2$S$_4$~\cite{FeCr2S4_Kalvius}), antiferromagnetic (e.g. MgV$_2$O$_4$~\cite{MgV2O4_Wheeler}), and incommensurate (e.g. CoCr$_2$O$_4$~\cite{CoCr2O4_ME_Yamasaki}), which is explained by the fact that both the A and B sublattices can incorporate magnetic ions. The B ions also form the so-called pyrochlore lattice, which is known to give rise to very strong geometrical frustration effects~\cite{Frustrated_Magnetism_Book}.

Detailed representation analysis of possible magnetic structures in spinels is given in~\cite{Izyumov_Spinels}. Most of the magnetic structures observed in spinels, especially those appearing at high temperatures, are described by the wave vector $\vec{k}=0$, i.e. the magnetic unit cell coincides with the crystal cell~\cite{Oles_Handbook}. Incommensurate magnetic structures are found in some spinels at temperatures below 20 -- 50~K and some of them are also shown to be ferroelectric (e.g. CoCr$_2$O$_4$~\cite{CoCr2O4_ME_Yamasaki} and ZnCr$_2$Se$_4$~\cite{ZnCr2Se4_Murakawa}). Therefore, despite the fact that spinels exhibit high temperature magnetic properties, ME effect in spinels is observed only at rather low temperatures. Here we show that chemical substitution in the A sublattice of spinels with sufficient degree of cationic ordering results in high temperature magnetically ordered phases becoming magnetoelectric.

The magnetic representation for the A and B positions in AB$_2$X$_4$ for $\vec{k}=0$ is given by~\cite{Izyumov_Spinels}
\begin{align}
d_M^A&=\rm{GM}^{4+}\oplus \rm{GM}^{5-},\nonumber\\
d_M^B&=\rm{GM}^{2+}\oplus \rm{GM}^{3+}\oplus 2\rm{GM}^{4+}\oplus \rm{GM}^{5+},\label{eq:Magnetic_Rep_B_cations}
\end{align}
respectively. The basis functions for IRs entering into the magnetic representations $d_M^A$ and $d_M^B$ are given in~\cite{Izyumov_Spinels}. It has to be noted, that since the spinel structure possesses spatial inversion $I$ a magnetic structure described by a single IR with $\vec{k}=0$ cannot induce electric polarization~\cite{Kovalev_SingleIR}. This is explained by the following. The symmetry of the paramagnetic phase is $G\otimes R$, where $G$ is the space group and $R$ is time inversion. When $\vec{k}=0$, for any of the IR's of $G\otimes R$ a unit matrix corresponds to either $I$ or $IR$. Therefore, upon a phase transition according to this IR one of these symmetry elements preserves in the ordered phase, but non of them allows non-zero electric polarization.

However, a magnetic phase transition with respect to IR GM$^{5-}$, which corresponds to appearance of a simple collinear antiferromagnetic ordering of spins of A cations, results in appearance of a linear ME effect. Denoting the antiferromagnetic ordering of A cation spins by $(L_x,L_y,L_z)$ the magnetoelectric interaction can be written in the form
\[
L_x(M_yP_z+M_zP_y)+L_y(M_zP_x+M_xP_z)+L_z(M_xP_y+M_yP_x),
\]
where $\vec{M}$ and $\vec{P}$ are magnetic moment and electric polarization, respectively. (Here and in the following we define the orthogonal $x$, $y$, and $z$ axes along the cubic edges.) Therefore, the magnetic structures with antiferromagnetically ordered spins of A cations possess linear ME effect. Such magnetic structures appear, for example, in MnAl$_2$O$_4$ below $T_N=42$~K~\cite{MnAl2O4_Krimmel}, Co$_3$O$_4$ below $T_N=40$~K~\cite{Co3O4_Roth}, and CoRh$_2$O$_4$ below $T_N=25$~K~\cite{CoB2O4_Suzuki}. The linear ME effect in A-site antiferromagnetic spinels has to be demonstrated experimentally yet.

The A-site antiferromagnetic structures in spinels, however, are rarely observed and occur at rather low temperatures. The magnetic phase transitions in spinels, and especially those taking place at high temperatures, more often occur with respect to IRs even under space inversion and entering into $d_M^B$. The resulting magnetic structures neither induce electric polarization nor allow ME effect, since they do not break inversion symmetry. However, cation substituted spinels A$'_{1/2}$A$''_{1/2}$B$_2$X$_4$ with partial or full ordering of A$'$ and A$''$ cations will possess magnetoelectric properties.

The $\vec{k}=0$ magnetic structures in spinels are most often described by IRs GM$^{4+}$ or GM$^{5+}$. The latter describes antiferromagnetic ordering of B cations. IR GM$^{4+}$ enters into both $d_M^A$ and $d_M^B$ and can induce ferromagnetic, ferrimagnetic or weak ferromagnetic structures, which, besides other causes, depends on whether both the A and B ions are magnetic or not.
We denote by $(f_x,f_y,f_z)$ and $(g_x,g_y,g_z)$ the magnetic order parameters transforming according to GM$^{4+}$ and describing ferromagnetic and antiferromagnetic ordering of B cations, respectively, whereas by $(a_1,a_2,a_3)$ the antiferromagnetic order parameter that transforms according to GM$^{5+}$.
The following ME interactions in spinels can be obtained
\begin{align}
s&(P_xf_yf_z+P_yf_zf_x+P_zf_xf_y),\label{eq:ME_invariants_ff}\\
s&(P_xg_yg_z+P_yg_zg_x+P_zg_xg_y),\label{eq:ME_invariants_gg}\\
s&(P_x(g_yf_z+g_zf_y)+P_y(g_zf_x+g_xf_z)+P_z(g_xf_y+g_yf_x)),\label{eq:ME_invariants_gf}\\
s&(P_xa_1a_3+P_ya_1a_2+P_za_2a_3),\label{eq:ME_invariants_aa}\\
s&(P_x(a_1g_y-a_3g_z)+P_y(a_2g_z-a_1g_x)+P_z(a_3g_x-a_2g_y)),\label{eq:ME_invariants_ag}\\
s&(P_x(a_1f_y-a_3f_z)+P_y(a_2f_z-a_1f_x)+P_z(a_3f_x-a_2f_y)).\label{eq:ME_invariants_af}
\end{align}
It follows from (\ref{eq:ME_invariants_ff}) -- (\ref{eq:ME_invariants_af}) that nonzero A cation ordering ($s\neq0$) in A$'_{1/2}$A$''_{1/2}$B$_2$X$_4$ spinels results in the fact that the magnetically ordered states induced by IRs GM$^{4+}$ or GM$^{5+}$ possess linear ME effect, whereas all but the $(\eta,0,0)$ phase state induced by them become improper ferroelectric. Therefore, when interpreting the influence of the A cation order on magnetic phase transitions in spinels one has to include the terms (\ref{eq:ME_invariants_ff}) -- (\ref{eq:ME_invariants_af}) into the expansion of the thermodynamic potential. The ME coefficients are, thus, directly proportional to the degree of atomic ordering $s$.

\section{Magnetoelectric coupling\label{sec:ME_coupling}}

A microscopic model of ME interactions based on local noncentrosymmetric surroundings of magnetic ions was recently suggested~\cite{Sakhnenko_Microscopy}. In current work we use this model to estimate the ME coefficients in cation-ordered spinels. In cation-disordered spinels with the cubic Fd$\bar{3}$m structure the A cations are located in noncentrosymmetric tetragonal positions (8a) with local symmetry $T_d$, whereas the B cations are in positions (16d) with centrosymmetric rhombohedral symmetry $D_{3d}$. Therefore, according to the microscopic model~\cite{Sakhnenko_Microscopy} the A cations and the oxygen ions, whose surrounding is polar with symmetry $C_{3v}$, can contribute to the ME effect in spinels with the symmetry Fd$\bar{3}$m.

The ordering of A cations in A$'_{1/2}$A$''_{1/2}$B$_2$X$_4$ spinels results in disappearance of the inversion symmetry operation and lowering of the crystal lattice symmetry to F$\bar{4}$3m. In the tetrahedral structure the atoms A$'$, A$''$, B, and X are located in positions (4a), (4d), (16e), and (16e), respectively~\cite{LiGaCr2O4_LiInCr2O4_Okamoto}. Therefore, local symmetry around the B cations becomes polar $C_{3v}$ and the local electric dipole moments of these ions can contribute to the ME effect.

\begin{figure}
\includegraphics[width=8.6cm]{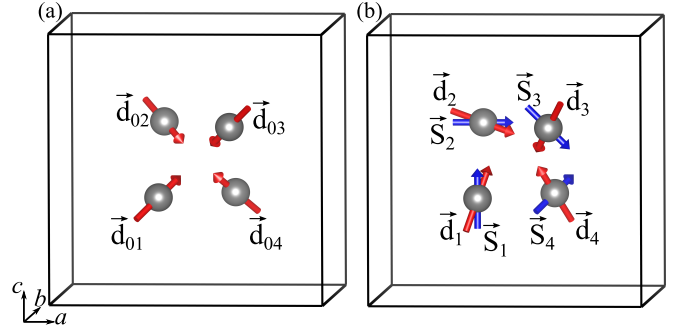}
\caption{\label{fig:Spinel_B_Atoms} (Color online) (a) Gray circles show the four atoms B$_i$ ($i=1,2,3,4$) present in the primitive cell of the cation ordered A$'_{1/2}$A$''_{1/2}$B$_2$X$_4$ spinel. Red arrows denote their respective electric dipole moments $\vec{d}_{0i}$ induced by local polar surroundings. (b) The same as (a) but taking the spins $\vec{S}_{i}$ (shown by blue arrows) into account. The spin-orbit coupling alters the electric dipole moments $\vec{d}_{0i}$ and result in spin-dependent electric dipole moments $\vec{d}_{i}$.}
\end{figure}

The primitive unit cell of the tetrahedral structure F$\bar{4}$3m contains four B cations B$_i$ ($i=1,2,3,4$) located in positions $(x,x,x)$, $(x,1-x,1-x)$, $(1-x,x,1-x)$, and $(1-x,1-x,x)$, respectively. Their respective electric dipole moments $\vec{d}_{0i}$ induced by local polar surroundings are equal in size and directed parallel to $[111]$, $[1\bar{1}\bar{1}]$, $[\bar{1}1\bar{1}]$, and $[\bar{1}\bar{1}1]$, respectively, as shown in Fig.~\ref{fig:Spinel_B_Atoms}(a). This ensures absence of macroscopic electric polarization ($\sum_i\vec{d}_{0i}=0$).

According to the microscopic model of ME interactions suggested earlier~\cite{Sakhnenko_Microscopy}, the spins $\vec{S}_i$ of the cations B$_i$ modify the electric dipole moments $\vec{d}_{0i}$ due to spin-orbit interaction as schematically shown in Fig.~\ref{fig:Spinel_B_Atoms}(b). This results in spin-dependent electric dipole moments $\vec{d}_{i}$ of the cations B$_i$, which may lead to nonzero macroscopic electric polarization $\vec{P}$ for certain spin configurations if $\sum_i\vec{d}_{i}\neq0$.

In order to build the microscopic model we closely follow the scheme developed in~\cite{Sakhnenko_Microscopy}. In the cubic Fd$\bar{3}$m structure the B cations are located in trigonally distorted oxygen octahedra. The local $D_{3d}$ symmetry splits the triply degenerate low lying $t_{2g}$ electron states into one $a_{1g}$ orbital and two degenerate $e'_g$ states. Therefore, for simplicity, as a zeroth order perturbation we consider the $a_{1g}$ state
\[
H_0|0\rangle=E_d|0\rangle,
\]
where $|0\rangle=|a_{1g}\rangle$, $H_0$ is the Hamiltonian including the crystal field of $D_{3d}$ symmetry and $E_d$ is the $a_{1g}$ energy level. In order to obtain spin-dependent electric dipole moments, we consider a single cation B$_1$, whereas the dipole moments of the remaining B cations can be obtained from B$_1$ by crystal symmetry operations. The A-site cation ordering reduces the crystal field symmetry around the B cations to $C_{3v}$, which is treated perturbatively. Compared to the $D_{3d}$ symmetry the $C_{3v}$ polar distortion gives additional contribution to the crystal field
\begin{equation}\label{EQ:VcrystalField}
V_{CF}=sc Z,
\end{equation}
where $c$ is coefficient and $s$ is included in order to reflect the fact that only $s\neq0$ results $V_{CF}\neq0$. In~(\ref{EQ:VcrystalField}) we consider only the lowest powers in crystal field coordinate expansion around the B cation. Here and in the following we define the orthogonal axes
\begin{align}
X&=\frac{1}{\sqrt{2}}(x-y),\nonumber\\
Y&=\frac{1}{\sqrt{6}}(x+y-2z),\nonumber\\
Z&=\frac{1}{\sqrt{3}}(x+y+z).\nonumber
\end{align}
The spin-orbit coupling is given by
\[
V_{SO}=-\lambda(\vec{L}\cdot \vec{S}),
\]
where $\vec{L}$ is the angular momentum operator, $\vec{S}$ is the spin, and $\lambda$ is the spin-orbit coupling constant. Thus, the perturbed hamiltonian has the form $H=H_0+V$, with $V=V_{CF}+V_{SO}$.

The perturbation $V$ mixes the unperturbed $a_{1g}$ state with other states and for simplicity it is sufficient to consider only the $4p$ states $H_0|p_\alpha\rangle=E_p|p_\alpha\rangle$ with the energy $E_p$, $\alpha=x,y,z$. One can write the perturbed eigenvector in the form
\[
|\psi\rangle=|0\rangle+\sum_\alpha A_\alpha |p_\alpha\rangle,
\]
where $A_\alpha$ are coefficients. The electric dipole moment is given then by
\[
\vec{d}=\langle\psi |e \vec{r}|\psi \rangle=\sum_\alpha A_\alpha
\langle 0|e\vec{r}|p_\alpha\rangle + c.c.,
\]
where $e$ is the electron charge. Performing the perturbations up to the third order we obtain the electric dipole moment of the B$_1$ cation as
\begin{align}
d_{X}&=\frac{d_0}{2}\frac{\lambda^2}{\Delta^2}S_XS_Z,\nonumber\\
d_{Y}&=\frac{d_0}{2}\frac{\lambda^2}{\Delta^2}S_YS_Z,\nonumber\\
d_{Z}&=d_{0}+d_{0}\frac{\lambda^2}{\Delta^2}(S^2_X+S^2_Y),\nonumber
\end{align}
with
\[
\vec{d}_{0}=2e\sum_\alpha\frac{V_{\alpha0} \langle0|\vec{r}|p_\alpha\rangle}{\Delta},
\]
where $\Delta=E_d-E_p$ and $V_{\alpha0}=\langle p_\alpha|V|0\rangle$. Thus, in addition to $\vec{d}_{0}$, which is the electric dipole moment induced by the local polar crystal field, the spin-orbit coupling gives rise to spin-dependent contribution to the electric dipole moment $\vec{d}$.

Obtaining similarly the remaining electric dipole moments of the cations B$_2$, B$_3$, and B$_4$, we find the macroscopic electric polarization $\vec{P}=\sum_i\vec{d}_i/v$ ($i=1,2,3,4$) as
\begin{align}
\begin{split}
P_{x}=q(&3(a_1g_y-a_3g_z)+g_zf_y+g_yf_z-2a_1a_3\\&+3(a_3f_z-a_1f_y)-4f_yf_z),
\end{split}\nonumber\\
\begin{split}
P_{y}=q(&3(a_2g_z-a_1g_x)+g_xf_z+g_zf_x-2a_1a_2\\&+3(a_1f_x-a_2f_z)-4f_xf_z),
\end{split}\label{eq:Polarization}\\
\begin{split}
P_{z}=q(&3(a_3g_x-a_2g_y)+g_yf_x+g_xf_y-2a_2a_3\\&+3(a_2f_y-a_3f_x)-4f_xf_y),
\end{split}\nonumber
\end{align}
where $v$ is the volume of the primitive cell, $q=d_0\lambda^2/(16\sqrt{3}v\Delta^2)$, and where we used the basis functions given in~\cite{Izyumov_Spinels} to rewrite the spins $\vec{S}_i$ of the cations B$_i$ in terms of the magnetic order parameters. The electric polarization~(\ref{eq:Polarization}), obtained from microscopic considerations, reflects the ME invariants~(\ref{eq:ME_invariants_ff}), ~(\ref{eq:ME_invariants_gf}), ~(\ref{eq:ME_invariants_aa}),~(\ref{eq:ME_invariants_ag}), and~(\ref{eq:ME_invariants_af}) and can be used to estimate the respective coefficients in the thermodynamic potential expansion. Performing the quantum perturbations to higher orders one can obtain additional contributions to polarization~(\ref{eq:Polarization}), which reflect in particular the existence of invariant~(\ref{eq:ME_invariants_gg}).

In order to estimate the electric polarization~(\ref{eq:Polarization}) we use the hydrogen-like orbitals and obtain $\langle0|\beta |p_\alpha\rangle\approx0.37a_0/Z$ ($\alpha,\beta=x,y,z$), where $a_0$ is the Bohr radius and $Z$ is the charge of the nucleus and core electrons in units of $e$. Taking Li$_{1/2}$Ga$_{1/2}$Cr$_2$O$_4$ as example~\cite{LiGaCr2O4_LiInCr2O4_Okamoto} we obtain $c\approx1.9\cdot 10^{-9}$~N and $v\approx140$~\AA. Using $Z\approx5$, $\lambda\approx0.05$~eV, $\Delta\approx1$~eV, $s=1$, and $f_\alpha\sim g_\alpha\sim a_i\sim1$ we find $d_0\sim 2\cdot10^{-31}$~C$\cdot$m and $q\sim0.13$~$\mu$C/m$^2$. The electric polarization $\vec{P}$ can, therefore, take values of the order of $0.1$ -- $0.5$~$\mu$C/m$^2$.

\section{Phenomenological models\label{sec:Phenomenology}}

\subsection{Ferromagnetic ordering\label{sec:GM4+}}

In this section we study ferromagnetic ordering in cation-ordered spinels, which is described by IR GM$^{4+}$. The thermodynamic potential expansion can be written in the form
\begin{multline}
\Phi=\frac{A_f}{2}I_1+\frac{b_1}{4}I_2+\frac{b_2}{4}I_1^2+\kappa I_{ME}+\frac{A_p}{2}I_P\\
-(f_xH_x+f_yH_y+f_zH_z),\label{eq:PotentialFeromagnetic}
\end{multline}
where $A_f$, $b_1$, $b_2$, $A_p$, and $\kappa$ are coefficients, $\vec{H}$ is magnetic field, $I_1=f_x^2+f_y^2+f_z^2$, $I_2=f_x^4+f_y^4+f_z^4$, $I_{ME}=s(P_xf_yf_z+P_yf_zf_x+P_zf_xf_y)$, and $I_p=P_x^2+P_y^2+P_z^2$. Following the usual premise of the phenomenological theory we assume $A_f=a_f(T-T_c)$, where $T$ is temperature, $T_c$ is the Curie temperature, and $a_f$ is a coefficient independent of $T$. Since the system is far from a proper ferroelectric phase transition we take $A_p\gg 0$.

In the paramagnetic and paraelectric phase $A_f>0$, $f_x=f_y=f_z=0$, $P_x=P_y=P_z=0$, and the magnetic susceptibility
\[
\chi_{\alpha\beta}=\frac{\partial f_\alpha}{\partial H_\beta},
\]
where $\alpha,\beta=x,y,z$, has the form $\chi_{\alpha\alpha}=1/A_f$ with other components equal to zero. The linear magnetoelectric coefficient
\[
\Lambda_{\alpha\beta}=\frac{\partial P_\alpha}{\partial H_\beta}
\]
is also zero. The potential~(\ref{eq:PotentialFeromagnetic}) allows two ferromagnetic phases, either of which may occur at $T=T_c$: an improper ferroelectric phase with $f_x=f_y=f_z\neq0$ and a paraelectric phase with $f_x=f_y=0$ and $f_z\neq0$.

\subsubsection{The phase $f_x=f_y=f_z\neq0$.}

\begin{figure}
\includegraphics[width=8.6cm]{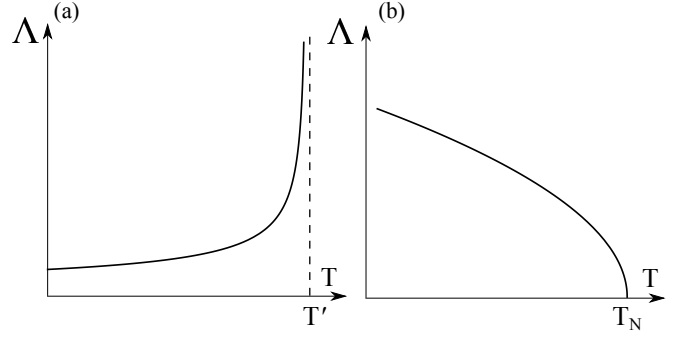}
\caption{\label{fig:Lambda} (a) Schematical temperature dependence of nonzero ME coefficients $\Lambda_{\alpha\beta}$ close to $T'$ ($T'=T_c$ or $T_N$) for the magnetic phase transitions described by IR GM$^{4+}$. (b) The same as (a) but for IR GM$^{5+}$.}
\end{figure}

In this phase (phase I), which is stable for $b_1>0$, the order parameters are given by $f_x=f_y=f_z=f$ and
\begin{equation}
P_x=P_y=P_z=-\frac{s\kappa}{A_p}f^2\label{eq:Polarization_from_f_squared}
\end{equation}
with
\[
f^2=-\frac{A_fA_p}{A_p(b_1+3b_2)-2s^2\kappa^2}.
\]
The magnetic susceptibility and magnetoelectric tensors represented in the $(X,Y,Z)$ coordinates are given by
\[
\chi_{XX}=\chi_{YY}=-\frac{A_p(b_1+3b_2)-2s^2\kappa^2}{2A_f(A_pb_1+s^2\kappa^2)},\qquad \chi_{ZZ}=-\frac{1}{2A_f},
\]
\[
\Lambda=\left(
\begin{array}{ccc}
\frac{s\kappa}{2f(A_pb_1+s^2\kappa^2)}&0&0\\
0& \frac{s\kappa}{2f(A_pb_1+s^2\kappa^2)}  &0\\
0&0&\frac{s\kappa f}{A_fA_p}
\end{array}
\right),
\]
respectively. Nondiagonal components of $\chi$ equal zero. Therefore, the ME tensor components are proportional to the degree of atomic ordering $s$ and diverge as $1/\sqrt{T_c-T}$ at $T<T_c$ as shown in Fig.~\ref{fig:Lambda}(a) since $f^2\sim(T_c-T)$.

\subsubsection{The phase $f_x=f_y=0$, $f_z\neq0$.}

In this phase (phase II), which is stable for $b_1<0$, the order parameters are given by $f_x=f_y=0$, $f_z=f$ and $P_x=P_y=P_z=0$ with
\begin{equation}
f^2=-\frac{A_f}{b_1+b_2}.\label{eq:f_phase_II}
\end{equation}
The magnetic susceptibility and magnetoelectric tensors represented in the $(x,y,z)$ coordinates are given by
\[
\chi_{xx}=\chi_{yy}=\frac{A_p(b_1+b_2)}{A_f(A_pb_1+s^2\kappa^2)},\qquad \chi_{zz}=-\frac{1}{2A_f},
\]
and
\begin{equation}
\Lambda=\left(
\begin{array}{ccc}
0&       \frac{s\kappa}{f(A_pb_1+s^2\kappa^2)}     &0\\
\frac{s\kappa}{f(A_pb_1+s^2\kappa^2)}   & 0  &0\\
0&0&0
\end{array}
\right),\label{eq:ME_tensor_phase_II}
\end{equation}
respectively. Nondiagonal components of $\chi$ equal zero. Similar to the previous case the ME tensor components are proportional to the degree of atomic ordering $s$ and diverge as $1/\sqrt{T_c-T}$ at $T<T_c$ as shown in Fig.~\ref{fig:Lambda}(a).

\subsubsection{Estimation of ME effect.\label{sec:GM4+_estimation}}

To make an order of magnitude estimation of the ME coefficient in the paraelectric phase II we proceed in the following way. We assume that the order parameter $f_\alpha$ in~(\ref{eq:PotentialFeromagnetic}) represents the magnetic moment of the B cation measured in Bohr magnetons $\mu_B$. It is convenient to introduce three constants
\begin{align}
c_1&=\frac{a_f}{b_1+b_2},\nonumber\\
c_2&=-\frac{2A_p(b_1+b_2)}{b_1A_p+\kappa^2},\nonumber
\end{align}
and
\[
c_3=\frac{\kappa}{A_p}.
\]
The first of them determines the magnetic moment $f^2=-c_1(T-T_c)$ in~(\ref{eq:f_phase_II}) and can be estimated as $c_1=0.25$~$\mu_B^2/K$. The second constant represents the ratio of susceptibilities $c_2=\chi_\perp/\chi_\parallel$, which we estimate as $c_2=1/2$. The third constant $c_3$ can be estimated from~(\ref{eq:Polarization_from_f_squared}) using the results of section~\ref{sec:ME_coupling}. Assuming that $f=5$~$\mu_B$ induces improper polarization of $0.5$~$\mu$C/m$^2$ we obtain $c_3=-0.02$~$\mu$C/($\mu_B$m)$^2$ (here and in the following we assume $s=1$). The phenomenological constant $a_f$, which is related to the paramagnetic susceptibility as $\chi^{-1}=a_f(T-T_c)$ can be estimated as $a_f=9.9\cdot 10^{27}$~m$^{-3}\cdot$K$^{-1}$, where we used the molar susceptibility $\chi_{mol}=C/(T-\Theta)$ of ferromagnetic spinel CuCr$_2$Te$_4$ with $C=4.28$~K$\cdot$emu/mol-f.u.~\cite{CuCr2Te4_Suzuyama}. The components of the ME tensor~(\ref{eq:ME_tensor_phase_II}) can then be expressed in the form
\[
\Lambda_{xy}=\Lambda_{yx}=\frac{c_2c_3\sqrt{c_1}}{2a_f\sqrt{T_c-T}}=-\gamma\sqrt{\frac{K}{T_c-T}},
\]
where $\gamma=2.17\cdot10^{-6}$~$\mu$C/(m$^2\cdot$Oe)$=0.027$~ps/m, which is about two orders of magnitude smaller than the maximum ME coefficient observed in Cr$_2$O$_3$~\cite{Cr2O3_Wiegelmann}. Assuming a dielectric constant of $\varepsilon=5.5$ observed in CoCr$_2$O$_4$~\cite{CoCr2O4_Lawes} we obtain $\gamma=0.45$~mV/(cm$\cdot$Oe), which is comparable to that of some BaTiO$_3$-based bulk particulate magnetoelectric composites~\cite{Composites_Vaz}.

\subsection{Antiferromagnetic ordering described by GM$^{4+}$\label{sec:GM4+_Antiferromagnetic}}

In this section we study antiferromagnetic ordering of B cations described by IR GM$^{4+}$. The thermodynamic potential expansion can be written in the form
\begin{multline}
\Phi=\frac{A_f}{2}I_1+\frac{A_g}{2}I_{g1}+\frac{b_{g1}}{4}I_{g2}+\frac{b_{g2}}{4}I_{g1}^2+wJ\\
+\kappa_1 I_{ME1}+\frac{A_p}{2}I_P-(f_xH_x+f_yH_y+f_zH_z),\label{eq:PotentialAntiFeromagnetic_GM4p}
\end{multline}
where $A_g$, $b_{g1}$, $b_{g2}$, $w$, and $\kappa_1$ are coefficients, $I_{g1}=g_x^2+g_y^2+g_z^2$, $I_{g2}=g_x^4+g_y^4+g_z^4$, $I_{ME1}=s(P_x(g_yf_z+g_zf_y)+P_y(g_zf_x+g_xf_z)+P_z(g_xf_y+g_yf_x))$, and $J=g_xf_x+g_yf_y+g_zf_z$. As discussed in section~\ref{sec:ME_coupling} the ME interaction~(\ref{eq:ME_invariants_gg}) appears in~(\ref{eq:Polarization}) only upon perturbations to higher orders and can be considered smaller than the other ME interactions. Therefore, we do not include this term in the expansion~(\ref{eq:PotentialAntiFeromagnetic_GM4p}).  Since the system is far from both the pure ferromagnetic and ferroelectric phase transitions we assume $A_f>0$ and $A_p>0$. Minimization of potential~(\ref{eq:PotentialAntiFeromagnetic_GM4p}) shows that the paramagnetic phase loses stability at $A_g=A_g^c=w^2/A_f$ and experiences a phase transition either to the ferroelectric phase (phase I) with  $g_x=g_y=g_z\neq0$, $f_x=f_y=f_z\neq0$, and $P_x=P_y=P_z\neq0$ or paraelectric phase (phase II) with $g_x=g_y=0$, $g_z\neq0$, $f_x=f_y=0$, $f_z\neq0$, and $P_x=P_y=P_z=0$. Here we consider only the latter phase for simplicity. Minimization of potential~(\ref{eq:PotentialAntiFeromagnetic_GM4p}) yields the order parameters
\begin{align}
f_z&=-\frac{w}{A_f}g_z,\label{eq:AntiFerro_WeakFerroMoment}\\
g_z^2&=-\frac{A_gA_f-w^2}{A_f(b_{g1}+b_{g2})}.\nonumber
\end{align}
The antiferromagnetic phase transition according to IR GM$^{4+}$ is a quasiproper ferromagnetic transition since the ferromagnetic moment transforms according to the same IR. The weak ferromagnetic moment arising due to~(\ref{eq:AntiFerro_WeakFerroMoment}) is usually about two orders of magnitudes smaller than the antiferromagnetic one. Therefore, the ratio $w/A_f$ can be estimated as 0.01 -- 0.05 and will be used as a small parameter in expansions.

Similar to the above case of a ferromagnetic phase transition we assume $A_g=a_g(T-T^\star)$ with $T^\star=T_N-w^2/(A_fa_g)$, where $T_N$ is the N\'{e}el temperature. In the paramagnetic phase at $A_g>A_g^c$ the nonzero components of the magnetic susceptibility tensor are given by
\[
\chi_{\alpha\alpha}=\frac{A_g}{A_fA_g-w^2}=\frac{1}{A_f}+\frac{w^2}{A_f^2a_g(T-T_N)},
\]
whereas the magnetoelectric tensor $\Lambda$ is equal to zero. Therefore, $\chi_{\alpha\alpha}$ diverges at $T=T_N$. However, due to the smallness of $w/A_f$ the temperature region of high $\chi_{\alpha\alpha}$ is very narrow.

In the phase II at temperatures below $T_N$ the nonzero components of magnetic susceptibility tensor are given by
\[
\chi_{xx}=\chi_{xx}=-\frac{w^2(b_{g1}+b_{g2})}{A_f^2a_gb_{g1}(T_N-T)},\quad \chi_{zz}=\frac{1}{A_f}+\frac{w^2}{2A_f^2a_g(T_N-T)},
\]
where for $\chi_{xx}$ and $\chi_{yy}$ we consider only the first term in their expansions with respect to $w/A_f$.

At temperatures close to $T_N$ the magnetoelectric tensor is given by
\begin{equation}
\Lambda=\left(
\begin{array}{ccc}
0&       \frac{2s\kappa_1w^2}{A_pb_{g1}A_f^2g_z}     &0\\
\frac{2s\kappa_1w^2}{A_pb_{g1}A_f^2g_z}   & 0  &0\\
0&0&0
\end{array}
\right),\label{eq:ME_tensor_Anti_g_phase_II}
\end{equation}
i.e. diverges at $T_N$ as $1/\sqrt{T_N-T}$ as shown in Fig.~\ref{fig:Lambda}(a), since
\[
g_z^2=\frac{a_g(T_N-T)}{b_{g1}+b_{g2}}.
\]
At temperatures significantly lower than $T_N$ we obtain
\begin{equation}
\Lambda=\left(
\begin{array}{ccc}
0&       -\frac{g_zs\kappa_1}{A_fA_p-g_z^2s^2\kappa_1^2}     &0\\
-\frac{g_zs\kappa_1}{A_fA_p-g_z^2s^2\kappa_1^2}   & 0  &0\\
0&0&0
\end{array}
\right).\label{eq:ME_tensor_Anti_g_phase_II_low_T}
\end{equation}
Here we again considered only the leading term in expansion of $\Lambda$ with respect to $w/A_f$.

To estimate the ME effect we introduce two constants
\[
c_1=\frac{a_g}{b_{g1}+b_{g2}},\qquad c_2=-\frac{a_gb_{g1}}{b_{g1}+b_{g2}}.
\]
The first of them determines the temperature dependence of the order parameter as $g_z^2=c_1(T_N-T)$, whereas $c_2$ reflects the temperature dependence of magnetic susceptibility as $\chi_{xx}=\chi_{yy}=w^2/(A_f^2c_2(T_N-T))$. The ME coefficients in~(\ref{eq:ME_tensor_Anti_g_phase_II}) take then the form
\[
\Lambda_{xy}=\Lambda_{yx}=\frac{2\sqrt{c_1}sw^2\kappa_1}{c_2A_f^2A_p\sqrt{T_N-T}}.
\]
Similarly to the case of section~\ref{sec:GM4+_estimation} we can use the estimates $c_1=0.25$~$\mu_B^2/K$ and $\kappa_1/A_p=0.02~\mu$C/($\mu_B$m)$^2$, whereas $c_2$ can be tentatively taken equal to $a_f=9.9\cdot 10^{27}$~m$^{-3}\cdot$K$^{-1}$ from section~\ref{sec:GM4+_estimation} (both $c_2$ in the current case and $a_f$ in section~\ref{sec:GM4+_estimation} determine the temperature dependence of magnetic susceptibility, whereas in the current case the fact that the weak ferromagnetism is considered is accounted for by the factor $(w/A_f)^2$ in the expression for $\chi_{xx}$ and $\chi_{yy}$). Assuming $w/A_f=0.05$ we obtain
\[
\Lambda_{xy}=\Lambda_{yx}=\gamma\sqrt{\frac{K}{T_N-T}},
\]
where $\gamma=5.4\cdot 10^{-4}$~ps/m, which is about two orders of magnitude smaller than that in section~\ref{sec:GM4+}.

To estimate the ME coefficients in~(\ref{eq:ME_tensor_Anti_g_phase_II_low_T}) we assume $A_fA_p\gg g_z^2s^2\kappa_1^2$ since the system is far from ferroelectric and ferromagnetic phase transitions. Therefore, nonzero coefficients in~(\ref{eq:ME_tensor_Anti_g_phase_II_low_T}) can be written as
$\Lambda_{xy}=\Lambda_{yx}=-g_zs\kappa_1/A_fA_p$. From the paramagnetic susceptibility $\chi=5\cdot10^{-3}$~$\mu_B$/T of Cr spins in LiGaCr$_4$O$_8$~\cite{LiGaCr2O4_LiInCr2O4_Okamoto} we can estimate $A_f$ as $A_f=\chi^{-1}$. Taking $g_z=1~\mu_B$ we obtain the components of ME tensor~(\ref{eq:ME_tensor_Anti_g_phase_II_low_T}) of the order of 1.3$\cdot 10^{-4}$~ps/m.

\subsection{Antiferromagnetic ordering described by GM$^{5+}$}

In this section we study antiferromagnetic ordering of B cations described by IR GM$^{5+}$. The thermodynamic potential expansion can be written in the form
\begin{multline}
\Phi=\frac{A_f}{2}I_1+\frac{A_a}{2}I_{a1}+\frac{b_{a1}}{4}I_{a2}+\frac{b_{a2}}{4}I_{a1}^2+\kappa_2 I_{ME2}\\
+\kappa_3 I_{ME3}+\frac{A_p}{2}I_P-(f_xH_x+f_yH_y+f_zH_z),\label{eq:PotentialAntiFeromagnetic_GM5p}
\end{multline}
where $A_a$, $b_{a1}$, $b_{a2}$, and $\kappa_2$ are coefficients, $I_{a1}=a_1^2+a_2^2+a_3^2$, $I_{a2}=a_1^4+a_2^4+a_3^4$, $I_{ME2}=s(P_x(a_1f_y-a_3f_z)+P_y(a_2f_z-a_1f_x)+P_z(a_3f_x-a_2f_y))$, and $I_{ME3}=s(P_xa_1a_3+P_ya_1a_2+P_za_2a_3)$. In this case we again assume $A_f\gg0$ and $A_p\gg0$ since the system is far from both the pure ferromagnetic and ferroelectric phase transitions. In the paramagnetic phase at $A_a>0$ the nonzero magnetic susceptibility tensor components are given by $\chi_{\alpha\alpha}=1/A_f$, whereas the linear ME tensor $\Lambda$ is zero.

The thermodynamic potential~(\ref{eq:PotentialAntiFeromagnetic_GM5p}) allows two antiferromagnetic phases, either of which becomes stable for $A_a<0$: a ferroelectric phase with $f_x=f_y=f_z=0$, $P_x=P_y=P_z=P$, and $a_1=a_2=a_3=a$ (phase I) and a paraelectric phase with $f_x=f_y=f_z=0$, $P_x=P_y=P_z=0$, $a_1=a$, and $a_2=a_3=0$ (phase II). Phase I is stable for $b_{a1}\geq-s^2\kappa_3^2/A_p$, whereas phase II for $b_{a1}<-s^2\kappa_3^2/A_p$.

In the phase I the order parameters are given by
\[
P=-\frac{s\kappa_3}{A_p}a^2
\]
and
\[
a^2=-\frac{A_a}{b_{a1}+3b_{a2}}.
\]
The magnetic susceptibility and ME tensors represented in the $(X,Y,Z)$ coordinates take the forms (here and in the following we do not consider the term in~(\ref{eq:PotentialAntiFeromagnetic_GM5p}) proportional to $\kappa_3$, since it gives minor contribution compared to the other terms)
\[
\chi_{XX}=\chi_{YY}=\frac{A_p}{A_fA_p-3a^2s^2\kappa_2^2}    ,\qquad   \chi_{ZZ}=\frac{1}{A_f}
\]
and
\begin{equation}
\Lambda=\left(
\begin{array}{ccc}
0&-\frac{as\kappa_2\sqrt{3}}{A_fA_p-3a^2s^2\kappa_2^2}&0\\
\frac{as\kappa_2\sqrt{3}}{A_fA_p-3a^2s^2\kappa_2^2}  &0&0\\
0&0&0
\end{array}
\right),\label{eq:ME_tensor_GM5+_phase_I}
\end{equation}
respectively. Nondiagonal components of $\chi$ equal zero.

In the phase II the order parameter is given by
\[
a^2=-\frac{A_a}{b_{a1}+b_{a2}},
\]
whereas $\chi$ and $\Lambda$ take the forms
\[
\chi_{xx}=\chi_{yy}=\frac{A_p}{A_fA_p-a^2s^2\kappa_2^2}    ,\qquad   \chi_{zz}=\frac{1}{A_f}
\]
and
\begin{equation}
\Lambda=\left(
\begin{array}{ccc}
0&-\frac{as\kappa_2}{A_fA_p-a^2s^2\kappa_2^2}&0\\
\frac{as\kappa_2}{A_fA_p-a^2s^2\kappa_2^2}  &0&0\\
0&0&0
\end{array}
\right),\label{eq:ME_tensor_GM5+_phase_II}
\end{equation}
respectively. Nondiagonal components of $\chi$ equal zero.

At temperatures $T\lesssim T_N$ below the antiferromagnetic transition temperature $T_N$ the ME coefficients~(\ref{eq:ME_tensor_GM5+_phase_I}) and~(\ref{eq:ME_tensor_GM5+_phase_II}) grow as $\sqrt{T_N-T}$ with decreasing temperature as shown in Fig.~\ref{fig:Lambda}(b) since $a^2\sim A_a\sim(T-T_N)$. Similarly to the estimation of ME coefficients~(\ref{eq:ME_tensor_Anti_g_phase_II_low_T}) performed in section~\ref{sec:GM4+_Antiferromagnetic} we can estimate the ME tensor components $\Lambda_{xy}=-\Lambda_{yx}$ in~(\ref{eq:ME_tensor_GM5+_phase_I}) and~(\ref{eq:ME_tensor_GM5+_phase_II}) to be of the order of $10^{-4}$~ps/m.

\section{Discussion}

In this work we have studied the magnetoelectric properties of A$'_{1/2}$A$''_{1/2}$B$_2$X$_4$ spinels with nonzero degree $s$ of A$'$ and A$''$ cation order. Disordered A$'_{1/2}$A$''_{1/2}$B$_2$X$_4$ spinels possess cubic Fd$\bar{3}$m structure at high temperatures, whereas nonzero $s$ results in lowering of crystal symmetry to F$\bar{4}$3m and disappearance of inversion symmetry operation. Spinels often exhibit high temperature magnetic phase transitions, which are usually governed by magnetic B cations and result in magnetic structures with unit cells coinciding with the crystallographic ones, i.e. with $\vec{k}=0$. The $\vec{k}=0$ magnetic representation of B cations~(\ref{eq:Magnetic_Rep_B_cations}) consists of IRs even under space inversion, which do not allow ME effect as shown in section~\ref{sec:Magnetic_Phase_Transitions}. However, cation ordering results in disappearance of inversion operation and emergence of ME effect due to interactions~(\ref{eq:ME_invariants_ff}) --~(\ref{eq:ME_invariants_af}).

The microscopic model of magnetically induced electric polarization is suggested in section~\ref{sec:ME_coupling}. The model is based on the recently proposed mechanism of ME effect~\cite{Sakhnenko_Microscopy} and directly takes into account the disappearance of inversion symmetry at the B cation sites upon atomic ordering in spinels. It is shown that in cation-ordered spinels local electric dipole moments at B positions are spin-dependent and result in macroscopic electric polarization for certain spin configurations. The electric polarization~(\ref{eq:Polarization}) obtained from microscopic theory is in accordance with macroscopic interactions~(\ref{eq:ME_invariants_ff}) --~(\ref{eq:ME_invariants_af}). The obtained value of magnetically induced polarization of the order of $0.5$~$\mu$C/m$^2$ is comparable to that of magnetoelectric CoCr$_2$O$_4$ with incommensurate magnetic ordering~\cite{CoCr2O4_ME_Yamasaki}. The maximum possible electric polarization value depends on the cation ordering degree $s$ as well as on the degree of polar distortion reflected by the coefficient $c$ in~(\ref{EQ:VcrystalField}). Therefore, in order to obtain higher polarization one has to increase $|c|$, which strongly depends on the difference of ionic radii of the ordering A$'$ and A$''$ cations. The spinels with substantially different ionic radii of A$'$ and A$''$ cations should generally possess stronger tendency to atomic ordering as well as larger $|c|$.

Basic phenomenological models of magnetic phase transitions according to the main ordering patterns of B cation spins are given in section~\ref{sec:Phenomenology}. The three suggested models describe the magnetic ordering according to the three-dimensional representations entering into the magnetic representation~(\ref{eq:Magnetic_Rep_B_cations}) and representing: (i) ferromagnetic (IR GM$^{4+}$), (ii) antiferromagnetic with weak ferromagnetism (IR GM$^{4+}$), and (iii) purely antiferromagnetic (IR GM$^{5+}$) ordering patterns of B cation spins. The temperature dependencies of the magnetic susceptibility and magnetoelectric tensors are determined. It is found that in the first two cases described by IR GM$^{4+}$ the ME tensor components diverge as $\gamma/\sqrt{T'-T}$ below the phase transitions temperature $T'$ ($T'=T_c$ or $T_N$) as shown in Fig.~\ref{fig:Lambda}(a). The coefficient $\gamma$ is about two orders of magnitude larger in the purely ferromagnetic case (i), than in the case of a weak ferromagnetic phase transition (ii). In case (iii) the ME tensor components grow below $T_N$ as $\sqrt{T_N-T}$ with decreasing temperature as shown in Fig.~\ref{fig:Lambda}(b).

In accordance with the general results for the upper limit of the ME coefficient, which is given by $(\varepsilon\chi)^{1/2}$~\cite{Upper_Bound_ME_constant_Brown}, we find that due to the fact that $\chi$ diverges at $T_c$ the ME susceptibility is largest in the case of a ferromagnetic phase transition (i), for which we obtain the ME coefficient of the order of 0.01~ps/m. In case (ii) the ME constants also diverge at $T_N$ since $\chi$ diverges, but possess additional factor $(w/A_f)^2\sim2.5\cdot10^{-3}$ due to the fact that in this case the crystal is weakly ferromagnetic. In case (iii) the obtained values of ME coefficients are rather low (of the order of 10$^{-4}$~ps/m), because $\chi$ remains low at antiferromagnetic phase transitions. Therefore, from the point of view of practical applications ordered spinels with purely ferromagnetic (or ferrimagnetic) phase transition are of primary interest. To this class belong, for example, ferrimagnetic Fe$_{0.5}$Cu$_{0.5}$Cr$_2$S$_4$~\cite{FeCuCr2S4_Kurmaev} and ferromagnetic Cu$_{0.5}$In$_{0.5}$Cr$_2$Se$_4$~\cite{Cu0.5In0.5Cr2Se4_Yokoyama}. The former spinel, however, is semiconducting and displays colossal magnetoresistance~\cite{Fe0.5Cu0.5Cr2S4_Ramirez_Colossal_MR}.

The ME properties of A$'_{1/2}$A$''_{1/2}$B$_2$X$_4$ spinels claimed in this work are due to the presence of 1:1 cation order at the A site. Therefore, as a result the ME coefficients become proportional to the degree of atomic ordering $s$. Complete order of A$'$ and A$''$ cations, which we assumed in current work, corresponds to $s=1$ and results in maximal ME properties. However, different crystal preparation conditions or thermodynamic history of the sample may result in only partial atomic ordering, which will reduce the ME response, or even in inhomogenous cation ordering when the order parameter $s$ becomes spatially dependent. For the description of the latter case one will have to employ averaging techniques, such as the one performed in~\cite{Sakhnenko_2005}.

Besides ME response the appearance of ME properties in cation-ordered magnetic spinels A$'_{1/2}$A$''_{1/2}$B$_2$X$_4$ should be also observed, for example, in infrared studies. Compared to non-substituted spinels AB$_2$X$_4$, which possess four active infrared modes~\cite{IR_studies_III_Preudhomme}, cation-ordered spinels A$'_{1/2}$A$''_{1/2}$B$_2$X$_4$ have seven active modes~\cite{IR_studies_V_Preudhomme}, which belong to the same IR T$_2$ of the F$\bar{4}$3m space group. It has to be noted, that the infrared spectra of ordered and disordered Li$_{0.5}$Ga$_{0.5}$Cr$_2$O$_4$ possess no drastic differences, which may be due to the difficulty of completely disordering the sample and the presence of locally ordered regions~\cite{IR_studies_V_Preudhomme}.

The magnetic phase transitions in ordered spinels to the phase states $(f,f,f)$, $(g,g,g)$, and $(a,a,a)$ of the cases (i), (ii), and (iii), respectively, are improper ferroelectric. Therefore, the temperature dependencies of the frequencies of infrared active modes should experience a kink at the phase transition. In contrast, under applied magnetic field the magnetic phase transitions with respect to both GM$^{4+}$ and GM$^{5+}$ become proper ferroelectric as seen from ME interactions~(\ref{eq:ME_invariants_ff}), (\ref{eq:ME_invariants_gf}), and~(\ref{eq:ME_invariants_af}). Therefore, the frequencies of infrared active modes should experience strong temperature and magnetic field dependence close to these magnetic phase transitions.

The suggested model can be used to estimate the magnetoelectric response in all cation-ordered A$'_{1/2}$A$''_{1/2}$B$_2$X$_4$ magnetic spinels. The prominent examples of such spinels are Li$_{1/2}$Ga$_{1/2}$Cr$_2$O$_4$ and Li$_{1/2}$In$_{1/2}$Cr$_2$O$_4$~\cite{LiGaCr2O4_LiInCr2O4_Okamoto}, Cu$_{1/2}$In$_{1/2}$Cr$_2$S$_4$~\cite{CuCr2S4_based_Kesler}, Fe$_{1/2}$Cu$_{1/2}$Cr$_2$S$_4$~\cite{Fe0.5Cu0.5Cr2S4_Palmer,FeCuGaCr2S4_Aminov}, Li$_{1/2}$FeRh$_{3/2}$O$_4$~\cite{Li0.5FeRh1.5O4_Kang}, and $\lambda$-Li$_{0.5}$Mn$_{2}$O$_4$~\cite{Li0.5Mn2O4_Julien}. The model can be also applied to AB$_2$X$_4$ spinels with sufficient degree of inversion and 1:1 cation ordering at the tetrahedral site, such as, for example, in FeIn$_2$S$_4$~\cite{FeIn2S4_Hill}. In section~\ref{sec:Phenomenology} macroscopic magnetoelectric response calculation for such spinels experiencing magnetic ordering with $\vec{k}=0$ is presented, which is based on the microscopic model given in section~\ref{sec:ME_coupling}. For spinels with complex magnetic ordering, such as when, for example, the magnetic unit cell is a multiple of the crystallographic one or when an incommensurate magnetic structure appears (i.e., when $\vec{k}\neq0$) the phenomenological description should be rewritten accordingly, whereas the results of the microscopic approach can still be used~\cite{Sakhnenko_Microscopy}.

\section{Conclusions}

In summary, we theoretically show that cation-ordered A$'_{1/2}$A$''_{1/2}$B$_2$X$_4$ magnetic spinels should display magnetoelectric properties. The value of ME effect is estimated using recently proposed microscopic model based on spin-dependent electric dipole moments of ions located in noncentrosymmetric positions. Three phenomenological models describing various ferromagnetic and antiferromagnetic ordering patterns of B cation spins are build and the corresponding ME responses are calculated.

The author acknowledges the financial support by RFBR grants Nos. 11-02-00484-a and 12-02-31229-mol\_a.


\end{document}